\documentclass[a4paper,twoside,notoc,11pt]{JHEP3}

\usepackage{epsfig,multicol}

\newcommand\fverb{\setbox\pippobox=\hbox\bgroup\verb}
\newcommand\fverbdo{\egroup\medskip\noindent%
                            \fbox{\unhbox\pippobox}\ }
\newcommand\fverbit{\egroup\item[\fbox{\unhbox\pippobox}]}
\newbox\pippobox

\newcommand{\beq} {\begin{equation}}
\newcommand{\eeq} {\end{equation}}
\newcommand{\beqa} {\begin{eqnarray}}
\newcommand{\eeqa} {\end{eqnarray}}

\newcommand{\ie}{{\it i.e.}}

\newcommand{\cf}{{\it cf.}}

\newcommand{\ieps}{i\varepsilon}
\newcommand{\order}[1]{${\cal O}\left(#1 \right)$}
\newcommand{\morder}[1]{{\cal O}\left(#1 \right)}
\newcommand{\eq}[1]{(\ref{#1})}

\newcommand{\pvec}{{\bf p}}

\newcommand{\half}{\frac{1}{2}}

\newcommand{\halft}{{\textstyle \frac{1}{2}}}

\newcommand{\Slash}[1]{ \parbox[b]{0.6em}{$#1$} \hspace{-0.55em}
                                \parbox[b]{0.55em}{ \raisebox{-0.2ex}{$/$}}}

\title{Dressing the Quark with QCD Condensates}

\author{Paul Hoyer\thanks{Research supported in part by the
European Commission under contract HPRN-CT-2000-00130.}\\
            Department of Physical Sciences and Helsinki Institute of
            Physics\\
            POB 64, FIN-00014 University of Helsinki, Finland \\
            E-mail: \email{paul.hoyer@helsinki.fi}}
\author{St\'ephane Peign\'e\\
            LAPTH, Chemin de Bellevue BP 110, F-74941 Annecy-le-Vieux
            Cedex, France\\
            E-mail: \email{peigne@lapp.in2p3.fr}}
\preprint{HIP-2003-15/TH \\ LAPTH-976/03 \\ \hepph{0304010}} 

\abstract{A condensate of $\pvec = 0$ partons in the perturbative vacuum gives
rise to a term $\propto\delta^4(p)$ in the free PQCD propagators.
The leading condensate contribution to the quark propagator can be exactly
summed since there is a factor $\delta^4(p)$ associated to each loop.
We calculate the dressed quark propagator in the presence of either a
gluon or a quark condensate, for a number of colors $N \to \infty$.
The dressed quark propagator satisfies a Dyson-Schwinger type equation
which can be exactly solved within our framework.

In the case of a gluon condensate the dressed quark propagator
has no pole, hence quarks cannot appear in asymptotic
states, and moreover the DS equation has a solution which
spontaneously breaks chiral symmetry. We also calculate the dressed
quark-photon vertex and verify that the corresponding
Ward-Takahashi identity is satisfied, and that the dressed self-energy
correction to the photon propagator does not shift the
physical photon pole.}


\begin{document}

\section{Introduction}

In this paper we consider the contributions to the quark propagator
from a ${\pvec}=0$ condensate term \cite{ch} in the free gluon and
quark propagators,
\beqa
i D_{ab}^{\mu\nu}(p) &=& -\delta_{ab}\, g^{\mu\nu} \left[\frac{i}{p^2+\ieps}
      + \lambda_g^2 \, (2\pi)^4 \, \delta^4(p)\right]
\label{gpropmod} \\
i S_{AB}(p) &=& \delta_{AB}\left[\frac{i\Slash{p}}{p^2+\ieps}+
      \lambda_q^3 \, (2\pi)^4 \, \delta^4(p)\right]
\label{qpropmod}
\eeqa
Due to the $\delta^4(p)$ factors which constrain loop integrals, the
leading order contributions in the condensate parameters $\lambda_g$
and $\lambda_q$ arising from Feynman diagrams of arbitrary complexity
may be resummed. (We simplify the topology of the diagrams by
considering only contributions that are of leading order also in the
number of colors $N$). We show that the quark propagator thus dressed
satisfies a Dyson-Schwinger (DS) type equation (Figs.~2 and 7) which can
be solved exactly.
For $\lambda_g \neq 0$ the resulting
propagator has a branch cut instead of a pole at $p^2=0$. Hence the
quark does not propagate to
the {\em in} and {\em out} states at
$t = \pm\infty$. The DS
equation also has a non-perturbative, chiral symmetry breaking
solution which does not reduce to the free propagator in the
$\lambda_g^2 \to 0$ limit.

We calculate the dressed quark-photon vertex and verify
that the Ward-Takahashi identity is satisfied. This ensures that the
quark loop correction to the photon propagator maintains the
masslessness of the photon.

The condensate terms in Eqs.~\eq{gpropmod} and \eq{qpropmod} arise
when one starts the perturbative expansion from a free state which
already contains gluon or quark pairs. Consider  for simplicity the
free hamiltonian of a scalar field $\phi(x)$,
\beq \label{h0}
H_0(t) = \half\int\frac{d^3\pvec}{(2\pi)^3}\left[|\pi(t,\pvec)|^2 +
(\pvec^2+m^2) |\phi(t,\pvec)|^2 \right]
\eeq
where $\pi$ is the canonical momentum. $H_0$ is a sum (over
3-momentum ${\pvec}$) of uncoupled harmonic oscillator hamiltonians.
The ground state wave function is thus
\beq \label{psi0}
\Psi_0(t) \propto\exp\left[-\half
\int\frac{d^3\pvec}{(2\pi)^3}E_\pvec\, |\phi(t,\pvec)|^2 \right]
\eeq
where $E_\pvec=\sqrt{\pvec^2+m^2}$. Correspondingly, in a path
integral formulation of perturbation theory it is straightforward to
verify \cite{ph0} that imposing a wave function of the form \eq{psi0}
at $t=\pm T$ and then letting $T \to\infty$ one recovers the standard
Feynman propagator of the scalar field. On the other hand, if one
uses a gaussian wave function with a more general coefficient
$C(\pvec)$,
\beq \label{psiC}
\Psi_C(t) \propto\exp\left[-\half \int\frac{d^3\pvec}{(2\pi)^3}
C(\pvec) |\phi(t,\pvec)|^2 \right]
\eeq
then the {\em on-shell} part of the free propagator of momentum $p$
is modified by a term proportional to
\beq \label{condterm}
i\left[C(\pvec)-E_\pvec\right]\delta(p^0-E_\pvec).
\eeq
The usual Feynman rules for calculating higher order contributions
apply provided the `condensate'
term \eq{condterm} is always included in the free propagator.

The $\lambda_g,\ \lambda_q$ terms in the free gluon and quark
propagators similarly arise from gaussian wave functions of the form
\eq{psiC} at $t=\pm\infty$, with $m=0$ and $C(\pvec=0) \neq 0$.
Explicit Lorentz invariance is maintained since the wave function is
modified only at $\pvec=0$.

The modified gaussian wave function \eq{psiC} may be interpreted as
describing a coherent superposition (or condensate) of particle pairs
by expanding
the exponential,
\beq \label{pairs}
\exp\left[-\halft C(\pvec) |\phi|^2 \right] = \sum_n
\frac{(-1)^n}{2^n n!} \left[C(\pvec)
-E_\pvec\right]^n|\phi(t,\pvec)|^{2n} \exp\left[-\halft E_\pvec\,
|\phi(t,\pvec)|^2 \right]
\eeq
Analogously, the condensate terms in Eqs.~\eq{gpropmod} and
\eq{qpropmod} may be viewed as arising from the presence of
$\pvec = 0$ gluon and quark pairs in the asymptotic $in$ and
$out$ states.

Formally, a perturbative expansion around any state that has a
non-vanishing overlap with the true ground state is equally
justified. Thus to our knowledge there is no theoretical reason to
exclude the condensate terms in the free gluon and quark propagators
in a perturbative expansion of QCD. As we shall see, these terms
strongly modify long distance physics but give only higher-twist
corrections at short distance. The summation of leading condensate
contributions that we consider here provides a kind of `tree'
approximation, to which corrections corresponding to higher
powers of $g^2$ may be systematically added. This tree
approximation allows studies of analyticity and unitarity in a novel
setting, which may be closer to the confinement domain than standard
PQCD.

Our approach may be compared to that of the QCD sum rules \cite{svz},
where external $p=0$ lines connect to vacuum matrix elements of quark
and gluon operators. Assuming that the strong coupling freezes in the
long distance regime at a sufficiently small value to make the
perturbative expansion relevant \cite{yd},
we need only a single parameter $\lambda_g$ ($\lambda_q$) in the
gluon (quark) propagator to calculate any Green function. This avoids the
use of sum rules, which require assumptions concerning their
saturation and parametrizations of vacuum expectation values.

In section 2 we calculate the effects of
a quark condensate ($\lambda_g = 0,\ \lambda_q \neq 0$) on the
dressed quark propagator,
quark-photon vertex and photon self-energy. We simplify the topology of the
contributing diagrams by taking the limit of a large number of colors,
$N\to\infty$ with $g^2N$ fixed \cite{thooft}.
The quark condensate introduces a parameter $\mu_q$
with the dimension of mass,
\beq
\mu_q^3 = g^2 N \, \lambda_q^3
\label{muq}
\eeq
The full expansion of any Green function $G$ is then a double sum of the form
\beq \label{pertexpq}
G_q=\sum_{\ell=0}^\infty (g^2N)^\ell \sum_{n=0}^\infty C^q_{\ell,n}\ \mu_q^{3n}
\eeq
where the index $q$ on $G_q$ indicates the presence of a quark
condensate $(\lambda_q \neq 0)$. We calculate the complete sum over
$n$ for $\ell=0$. This is possible since for $\ell=0$ there is a
$\delta^4(p)$ factor from the free quark propagator \eq{qpropmod} in
each loop integral. The remaining sum over $\ell$ is then of usual
perturbative form, involving higher powers of the coupling $g^2N$ and
an increasing number of non-trivial loop integrals.

We neglect contributions where the condensate $p=0$ lines
are dressed, giving factors of $\delta^4(0)$ proportional to the
volume of space-time. Such terms should factorize from physical
quantities.
Similarly contributions where condensate lines interact with
themselves are dropped, being singular and of a non-local nature.

We find that the quark acquires an effective, `constituent' mass, as
expected from the fact that the inclusion of quark pairs in the
asymptotic states ($\lambda_q \neq 0$) breaks chiral symmetry \cite{ph0}.
Gauge invariance is
preserved and the photon remains massless.

In section 3 we study in a similar way the effects of a gluon
condensate ($\lambda_g \neq 0,\ \lambda_q=0$). The gluon condensate
mass parameter is
\beq
\mu_g^2 = g^2 N \, \lambda_g^2
\label{mug}
\eeq
and the full QCD expansion of any Green function is of the form
\beq \label{pertexpg}
G_g=\sum_{\ell=0}^\infty (g^2N)^\ell \sum_{n=0}^\infty C^g_{\ell,n}\ \mu_g^{2n}
\eeq
We evaluate the full $\ell=0$ `tree' contribution in the
$N \to\infty$ limit. The DS equation (Fig.~7) for the dressed quark
propagator is an ordinary second order algebraic equation.
The solution has a cut
instead of a pole at $p^2=0$, implying that the quark is removed from
the spectrum of on-shell states. Surprisingly, there is also a
non-perturbative solution which breaks chiral invariance. We verify
the Ward-Takahashi identity, which ensures the masslessness of the
photon within dimensional regularization.

We do not evaluate here the dressed gluon propagator, nor do we
consider the case of a simultaneous gluon and quark
condensate ($\lambda_g,\ \lambda_q \neq 0$).
The topological structure of the contributing Feynman diagrams
is more complicated in these cases, principally due to contributions
from the four-gluon vertex.
We have verified that the
gluon condensate contributions of order $\mu_g^2$ and $\mu_g^4$ (for $\ell=0$)
to the gluon self-energy have the required transverse structure.
We postpone to a future work the calculation of the `tree' ($\ell=0$)
dressed gluon propagator to all orders in the gluon condensate parameter
$\mu_g^2$. Section 4 contains an outlook.

\section{Quark condensate: $\lambda_g = 0$, $\lambda_q \neq 0$.}

\subsection{The dressed quark propagator}

The first correction to the quark propagator ($\ell=0,\ n=1$ in
\eq{pertexpq}) is given by the diagram of Fig.~1a, where only the
$\lambda_q^3$ term of the free propagator \eq{qpropmod} is kept. This
condensate term is proportional to $\delta^4(k)$, where $k$ is the
internal quark momentum, and is indicated pictorially by a cut line.
The quark self-energy, defined by amputating the external quark
lines, is~\cite{ph0}:
\beq
\label{sigmaq}
\Sigma_q(p) = \frac{2\mu_q^3}{p^2+\ieps}
\eeq
\EPSFIGURE{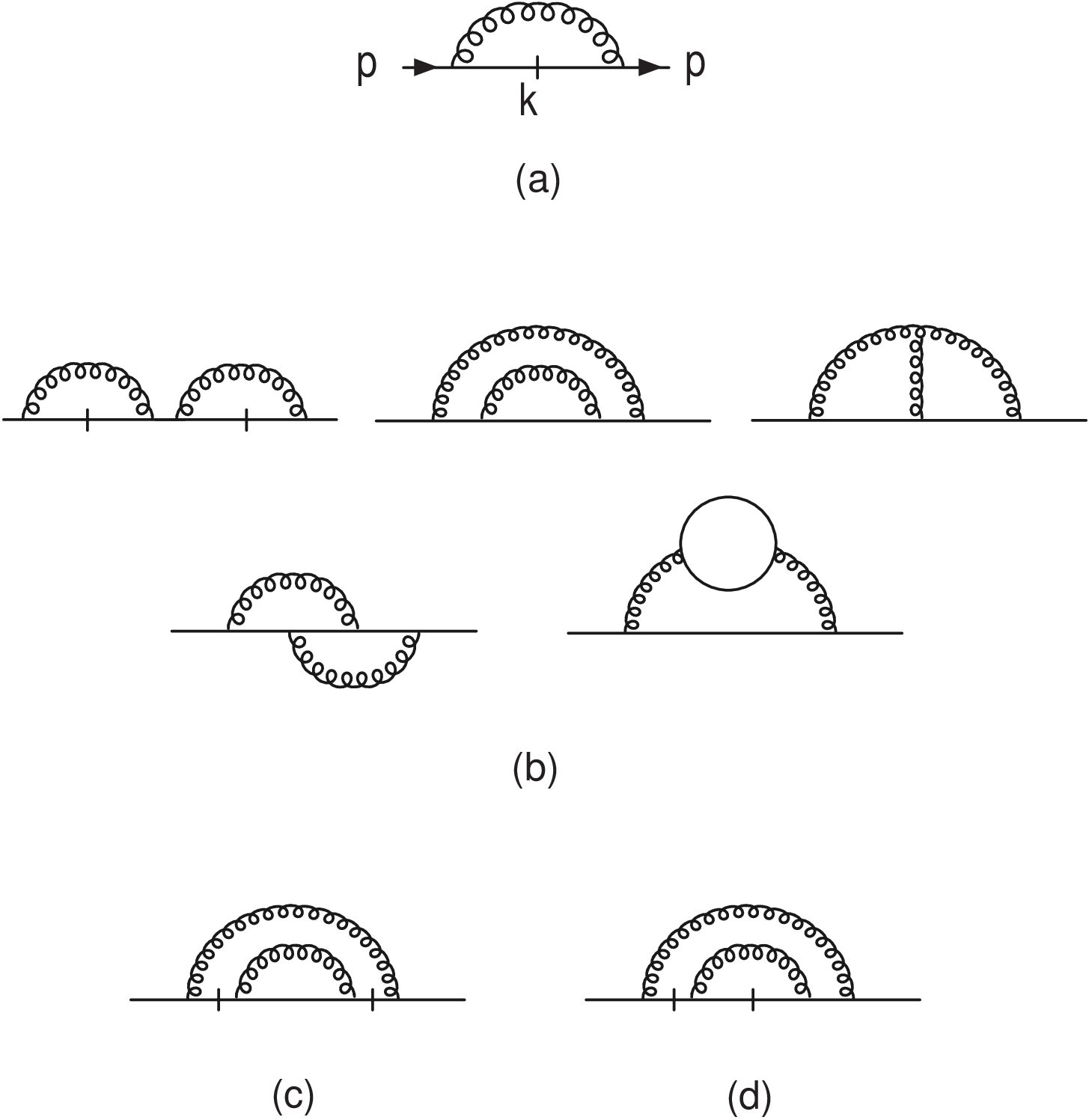,width=12cm}{(a) The quark self-energy $\Sigma_q$
at \order{\mu_q^3}. The cut internal quark line indicates that only
the term $\propto \delta^4(k)$ is kept for the internal propagator.
(b) Diagrams which might a priori contribute to order $\mu_q^6$.
(c) A contribution proportional to $\delta^4(0)$.
(d) Another singular contribution.}
 From now on the color indices and trivial color factors such as
$\delta_{AB}$ will be implicit. In \eq{sigmaq} we approximated $C_F =
(N^2-1)/2N$ by its large $N$ limit $C_F \simeq N/2$.

At the next order ($n=2$) the five Feynman diagrams of Fig.~1b, which
have at least two internal quark lines, may contribute. Of these, the
last two diagrams are suppressed in the large $N$ limit. The second
diagram can contribute only via the two possible cuts shown in
Figs.~1c and 1d. Fig.~1c is a self-interaction of a condensate line
and thus proportional to $\delta^4(0)$, \ie, to the volume $\int d^4
x$ of space-time. We expect such contributions to factorize from
connected Green functions and do not consider them further. Fig.~1d
involves scattering of condensate lines and is similarly non-local.
The third diagram of Fig.~1b can be dropped for the same reason.
Hence only the first diagram of Fig.~1b contributes at order
$\ell=0,\ n=2$.

The above analysis generalizes in a straightforward way to higher
powers of $\mu_q^3$, \ie, to terms in \eq{pertexpq} with $\ell=0$ and
arbitrary $n$. As a result, only diagrams corresponding to the
geometric series dress the quark propagator $S_q(p)$:
\beq
i S_q(p) = \frac{i}{\Slash{p}} + \frac{i}{\Slash{p}} (i \Sigma_q(p))
\frac{i}{\Slash{p}} + \frac{i}{\Slash{p}} (i \Sigma_q(p))
\frac{i}{\Slash{p}} (i \Sigma_q(p)) \frac{i}{\Slash{p}}  + \ldots
\eeq
implying the self-consistency equation shown in Fig.~2,
\beq \label{Sq0}
S_q(p) = \frac{1}{\Slash{p}} - \frac{1}{\Slash{p}} \Sigma_q(p) S_q(p)
\eeq
with solution
\beq \label{Sq}
S_q(p) = \frac{1}{\Slash{p} + \Sigma_q(p)} = \frac{1}{\Slash{p} + 2
      \mu_q^3/p^2} = \frac{p^4 \Slash{p} - 2\mu_q^3 p^2}{p^6 - (2\mu_q^3)^2}
\eeq
\EPSFIGURE[t]{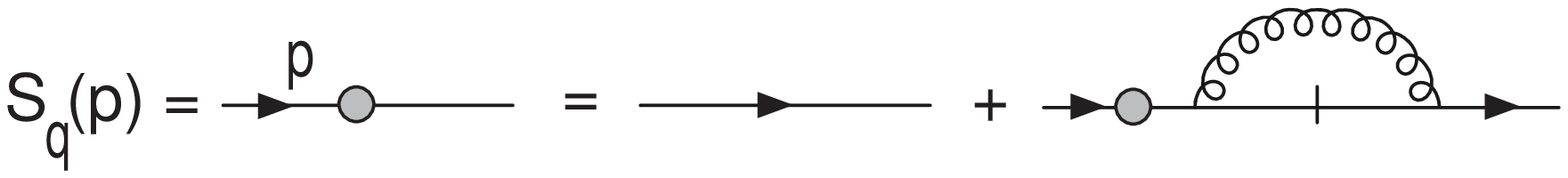,width=12cm}{Self-consistency equation for the dressed
   quark propagator $S_q(p)$. The shaded blob denotes the quark 
condensate dressing.}
The poles of $S_q(p)$ at $(p^2)^3 = (2 \mu_q^3)^2$ correspond to a
`constituent' quark mass
\beq \label{Mq}
M_q = \mu_q  \sqrt[3]{2}
\eeq
and two complex poles at $p^2 = M_q^2 \exp(\pm 2i\pi/3)$.
The appearance of a constituent quark mass $M_q \neq 0$ could be
anticipated from the fact that
the quark condensate gives a chirally non-invariant
free propagator \eq{qpropmod}.

We stress that the quark condensate contributions are power suppressed
at short distance,
\beq \label{asp2}
p^2 \to\infty \ \ \Rightarrow \ \
S_q(p) = \frac{1}{\Slash{p}} +\morder{\frac{\mu_q^3}{p^4}}
\eeq

\subsection{Dressed quark-photon vertex}

In the $N \to \infty$ limit, the leading behaviour in $\mu_q$ of the
quark-photon vertex $\Gamma_q^{\mu}(k, \bar k)$ is given by the series
shown in Fig.~3, where $\bar k = k - p$. This specific structure
arises because cutting both the quark and antiquark lines appearing
in between two
successive gluon exchanges would prevent the momentum $p$ from
flowing through the diagram from left to
right.
\EPSFIGURE{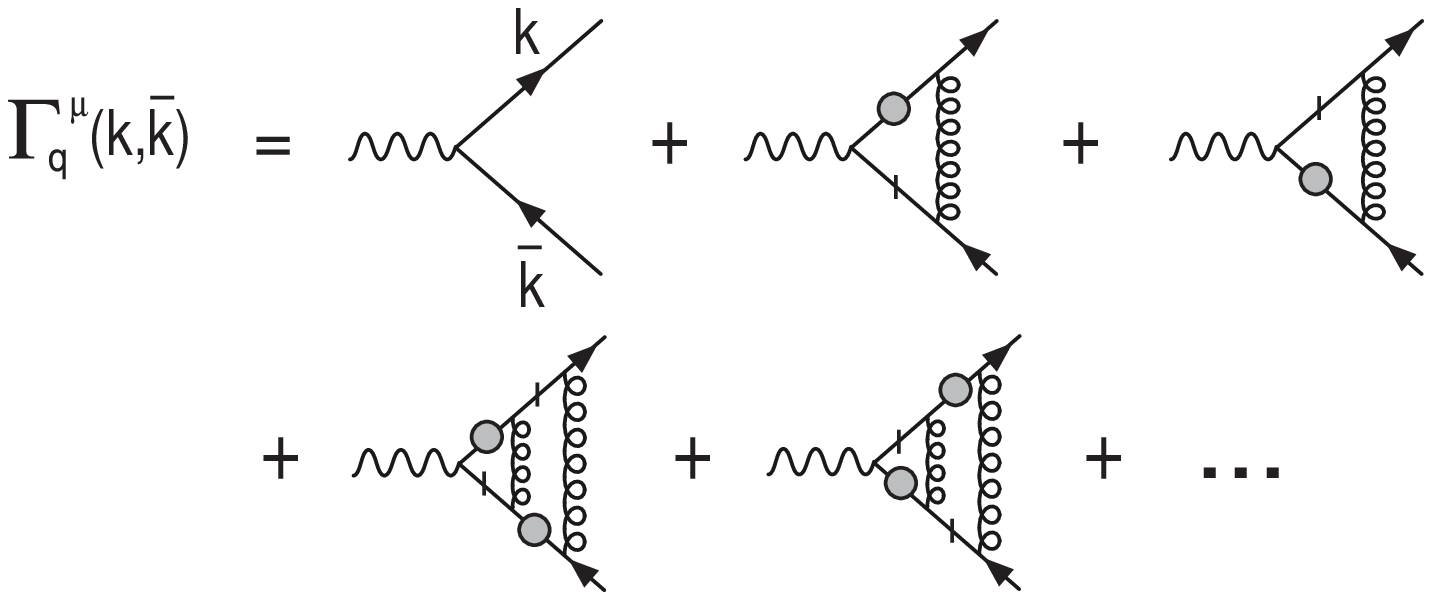,width=10cm}{Dressed quark-photon vertex
$\Gamma_q^{\mu}(k, \bar k)$. The shaded blob denotes the 
dressed quark propagator given in \eq{Sq} (and in Fig.~2).}
\EPSFIGURE{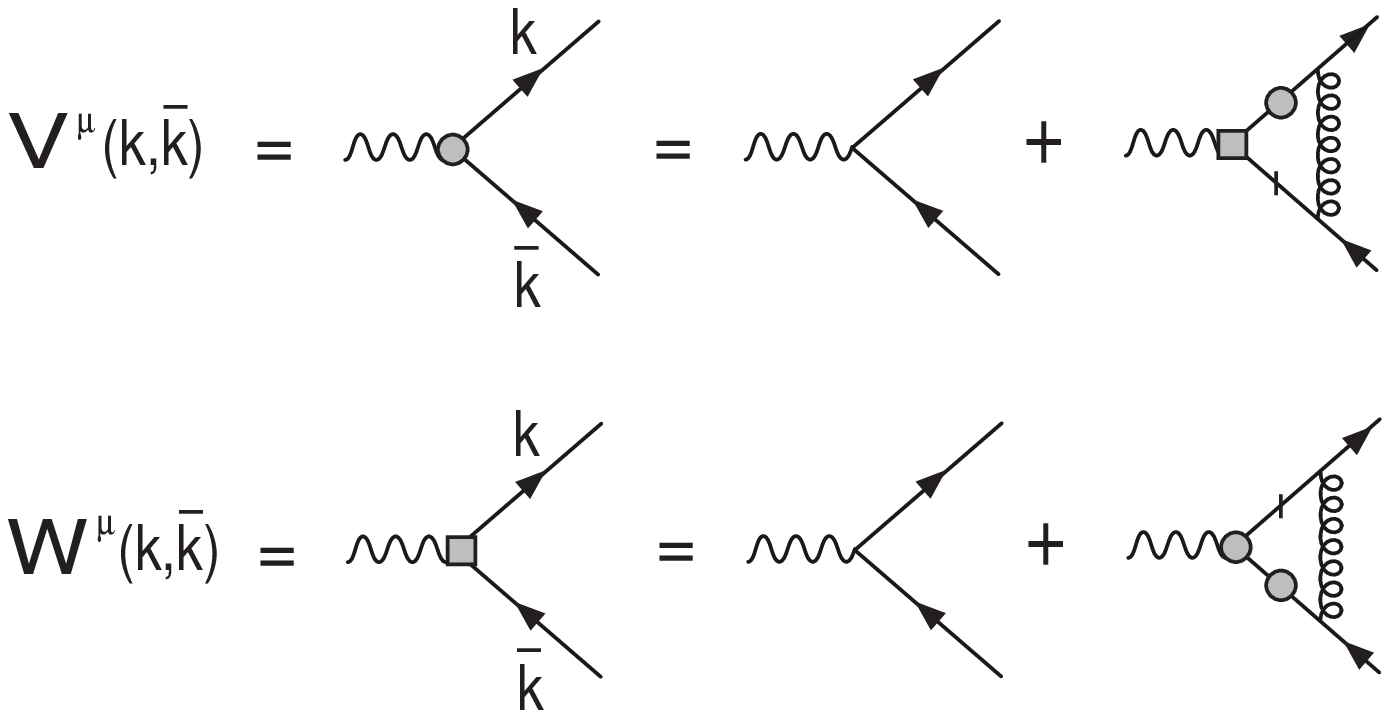,width=10cm}{Coupled equations for $V^{\mu}$ and
$W^{\mu}$.}
One can easily check that
\beq
\label{Gammaqmudef}
\Gamma_q^{\mu} = V^{\mu} + W^{\mu} - \gamma^{\mu}
\eeq
where the effective vertices $V^{\mu}$ and $W^{\mu}$ satisfy the
coupled equations shown in Fig.~4:
\beqa
\label{VWimplicit}
V^{\mu}(k, \bar k) &=& \gamma^{\mu} - \frac{\mu_q^3}{2 {\bar k}^2}
\gamma^{\nu} S_q(p) W^{\mu}(p, 0) \gamma_{\nu} \nonumber \\
W^{\mu}(k, \bar k) &=& \gamma^{\mu} - \frac{\mu_q^3}{2 {k}^2}
\gamma^{\nu} V^{\mu}(0, -p ) S_q(-p) \gamma_{\nu}
\eeqa
where $S_q(p)$ is given in \eq{Sq}.
The equations \eq{VWimplicit} are solved in Appendix A.
The result for $\Gamma_q^{\mu}$ given by \eq{Gammaqmudef} reads:
\beq
\label{Gammaqmu}
\Gamma_q^{\mu}(k, \bar k) = \gamma^{\mu} + \frac{2 \mu_q^3}{p^2} \left(
      \frac{1}{k^2} - \frac{1}{{\bar k}^2}\right) p^{\mu} -
\frac{2 \mu_q^6}{2 \mu_q^6 -p^6} \left( \frac{1}{k^2} +
      \frac{1}{{\bar k}^2}\right) \left( \Slash{p} p^{\mu} - p^2
\gamma^{\mu} \right)
\eeq
Using \eq{Sq} and \eq{Gammaqmu} one readily verifies the Ward-Takahashi
identity:
\beq
\label{WT}
p_{\mu} \Gamma^{\mu}(k, \bar k) = S(k)^{-1} -  S({\bar k})^{-1}
\eeq

\subsection{Photon self-energy}

The photon self-energy $\Pi_q^{\mu\nu}$ in the presence of a quark
condensate ($\lambda_q \neq 0$) can be expressed in terms of the
dressed quark propagator $S_q(p)$ and vertices $V^{\mu}$ and
$W^{\mu}$, as shown in Fig.~5. At leading order in $\mu_q$ only the
condensate term of the internal free quark propagator \eq{qpropmod}
is kept (as indicated by the cut line in Fig.~5). The loop integral
is then trivial, and
\beq
\label{Aqmunudef}
\Pi_q^{\mu\nu}(p)= e^2 N \lambda_q^3 \left\{ {\rm Tr} \left[ \gamma^{\nu}
        V^{\mu}(0, -p ) S(-p) \right]  + {\rm Tr} \left[ \gamma^{\nu} S(p)
        W^{\mu}(p, 0)  \right] \right\}
\eeq
\EPSFIGURE{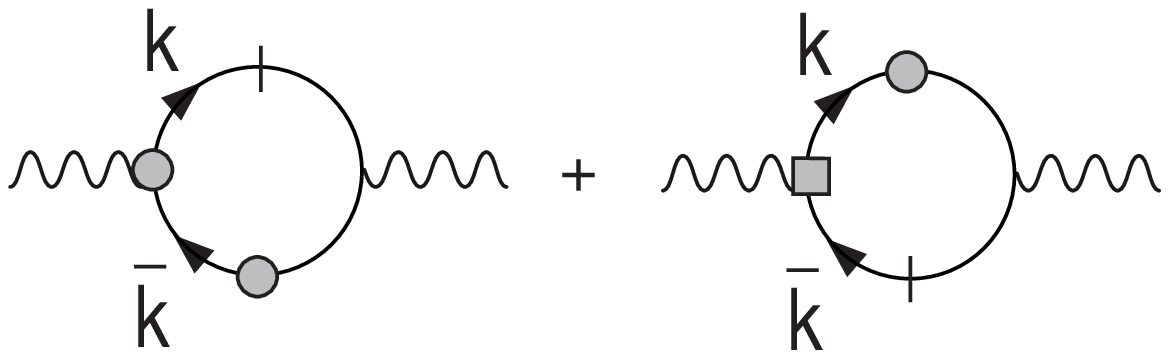,width=9cm}{Photon self-energy $\Pi_q^{\mu\nu}$
in a quark condensate.} 
Using the results of Appendix A for $V^{\mu}$ and  $W^{\mu}$ one has:
\beqa
V^{\mu}(0, -p) &=& \gamma^{\mu} - \frac{\mu_q^3}{2 p^2}
\left[A \gamma^{\mu} + B p^{\mu} + C \Slash{p} p^{\mu} \right] \nonumber \\
W^{\mu}(p, 0) &=& \gamma^{\mu} - \frac{\mu_q^3}{2 p^2}
\left[A \gamma^{\mu} - B p^{\mu} + C \Slash{p} p^{\mu} \right]
\eeqa
where $A$, $B$, $C$ are given in \eq{ABC}. A short calculation yields:
\beq
\Pi_q^{\mu\nu}(p)= 16 e^2 \frac{N \lambda_q^3 \mu_q^3}{2 \mu_q^6 - p^6} \,
(p^2  g^{\mu\nu} - p^{\mu}p^{\nu}) \equiv \Pi_q(p^2)\,(p^2  g^{\mu\nu}
- p^{\mu}p^{\nu})
\eeq
The fact that $\Pi_q(p^2=0)$ is finite implies that the photon
remains massless after the self-energy correction.

\section{Gluon condensate: $\lambda_g \neq 0$, $\lambda_q = 0$.}

\subsection{Dressed quark propagator}

At first order in $\mu_g^2$ ($\ell=0,\ n=1$ in \eq{pertexpg}), the
quark propagator is modified by the correction shown in Fig.~6a. The
internal gluon line is cut,
indicating that only the condensate term proportional to
$\lambda_g^2\, \delta^4(k)$ is kept in the free gluon propagator
\eq{gpropmod}. At second order, namely $\mu_g^4$ ($\ell=0,\ n=2$),
only the two diagrams shown in Fig.~6b contribute. Other diagrams,
such as the three last ones in Fig.~1b
(with two internal cut {\it gluon} instead of {\it quark} lines) can
be neglected for reasons similar to the case $\lambda_q \neq 0$,
$\lambda_g = 0$ studied in the previous
section. They are either suppressed by powers of $1/N$ in the large
$N$ limit, or represent non-local contributions which
should factorize from physical quantities.
\EPSFIGURE{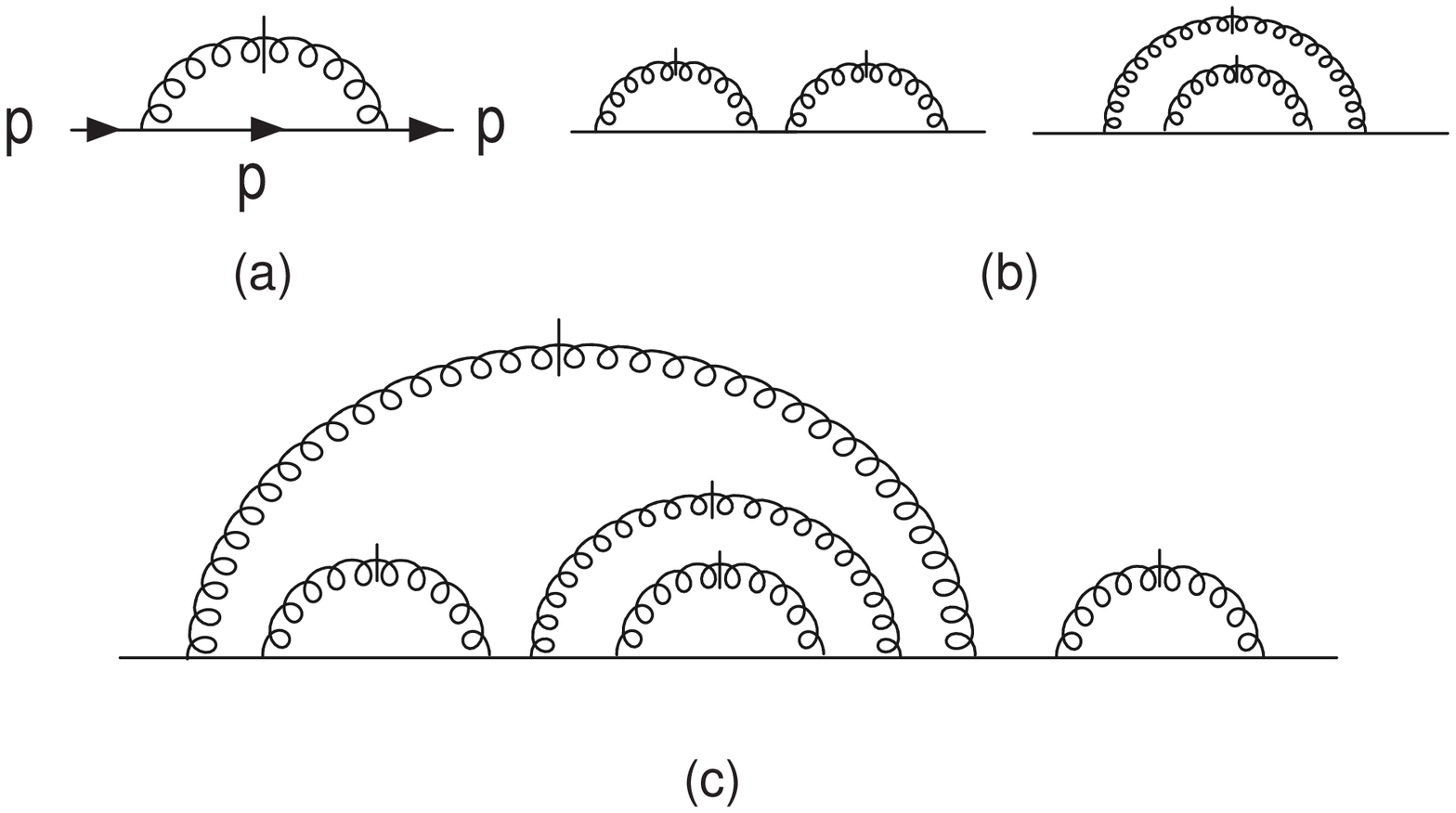,width=12cm}{Quark propagator $S_g(p)$
in a gluon condensate,
(a) at order $\mu_g^2$ and (b) at order $\mu_g^4$. (c) A generic diagram
of order $\mu_g^{2n}$.}
\EPSFIGURE{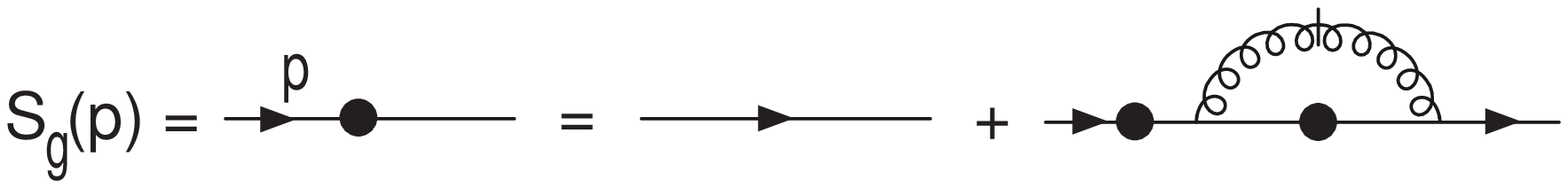,width=12cm}{Implicit equation for the
dressed quark propagator $S_g(p)$.
The full blob denotes the gluon condensate dressing.} 

At higher orders ($\ell=0,\ n>2$
in \eq{pertexpg}) the relevant diagrams contributing to the quark
propagator in the large $N$ limit are of the type shown in Fig.~6c,
where all internal gluon lines are cut. All loop integrals are
trivial due to the $\delta^4(k)$ factors from the condensate terms,
showing that we are in effect dealing with a tree approximation. Note
that we take into account one-particle irreducible as well as
reducible diagrams, thus including all $\ell=0$ contributions.

One may readily check that the complete set of relevant diagrams for
the quark propagator $S_g(p)$ is generated by iterating the implicit
equation shown in Fig.~7, which reads
\beq
\Slash{p} S_g(p) = 1 - \halft\mu_g^2\, \gamma^{\mu} S_g(p)
\gamma_{\mu} S_g(p)
\label{Sg0}
\eeq
where we used $C_F = N/2$ at leading order in $N$.

Lorentz invariance constrains the quark propagator to be of the form
\beq \label{genform}
S_g(p) = a(p^2)\,\Slash{p}+b(p^2)
\eeq
Substituting this in \eq{Sg0} one finds the equations
\beqa \label{condSg}
b(1+\mu_g^2 a) &=& 0 \nonumber\\
ap^2(1-\mu_g^2 a) &=& 1-2\mu_g^2 b^2
\eeqa

A chiral symmetry conserving quark propagator must have $b=0$. With
this constraint the second order equation for $a$ gives
\beq \label{csbcons}
S_{g1}(p) = \frac{2 \Slash{p}}{p^2 + \sqrt{p^2(p^2-4\mu_g^2)}}
\eeq
where we chose the sign of the square root so as to ensure that the
propagator approaches the free one in the $p^2 \to\infty$ limit:
\beq \label{asp2bis}
p^2 \to\infty \ \ \Rightarrow \ \
S_{g1}(p)=\frac{1}{\Slash{p}}\left[1+\morder{\frac{\mu_g^2}{p^2}}\right]
\eeq

It is important to note that the $p^2 =0$ quark pole of the
($\lambda_q=0$) free propagator \eq{qpropmod} has been removed by the dressing.
The dressed propagator $S_{g1}(p)$
instead has branch point singularities at $p^2 =0$ and
$p^2=4\mu_g^2$. In Appendix B (\cf\ Eq.~\eq{Sgasymp}) we show that
this removes the quark from the set of {\em in} and {\em out} states,
in the sense that the Fourier transformed propagator vanishes at
large times,
\beq \label{ast}
|S_{g1}(t, \vec{p})| \ \ \mathop{\sim}_{|t| \to \infty} \ \
\morder{1/\sqrt{|t|}}
\eeq
This raises interesting questions about analyticity and unitarity,
which we hope to return to in a future paper.

If one does not impose chiral symmetry and thus allows $b\neq 0$ in
\eq{condSg} one finds
\beq \label{csbbreak}
S_{g2}(p) = -\frac{1}{\mu_g^2}\left(\Slash{p} \pm \sqrt{p^2 +
\halft\mu_g^2}\right)
\eeq
This solution is singular for $\mu_g^2 \to 0$ and hence does not have
a power expansion in $\mu_g^2$ of the form \eq{pertexpg}. It emerges
as a `non-perturbative' solution of the implicit equation \eq{Sg0}. Like
the chiral symmetry conserving solution $S_{g1}$, the propagator
$S_{g2}$ has no quark pole, only a branch point at $p^2 = -\mu_g^2/2$.

The solution $S_{g2}$ does not approach the perturbative propagator
$1/\Slash{p}$ at large $p^2$, and must therefore be discarded on
physical grounds at short distance. However, we note that $S_{g2} =
S_{g1}$ at $p^2 = -\mu_g^2/2$. In a loop integral over $p^0$ it is
then possible (for $\pvec^2 > \mu_g^2/2$) to choose the $S_{g2}$
solution in the interval $-\sqrt{\pvec^2-\mu_g^2/2} \leq p^0 \leq
\sqrt{\pvec^2-\mu_g^2/2}$. This will give Green functions that break
chiral symmetry. It is interesting to find chiral symmetry breaking
without having introduced a quark condensate
in the asymptotic states ($\lambda_q=0$).

\subsection{Dressed quark-photon vertex}

The $q\bar q\gamma$ vertex $\Gamma_g^{\mu}(k, \bar k)$ is given by
the series shown in Fig.~8 for large $N$ and $\lambda_g \neq 0$.
Again, the loop integrals are trivially performed in this $\ell=0$
approximation of \eq{pertexpg} since all internal gluon lines are
cut. The infinite sum satisfies the implicit equation of Fig.~9,
\beq
\label{Gammagmuimplicit}
\Gamma_g^{\mu}(k, \bar k) =  \gamma^{\mu} - \halft\mu_g^2
\gamma^{\nu} S_g(k) \Gamma_g^{\mu}(k, \bar k) S_g(\bar k) \gamma_{\nu}
\eeq
\EPSFIGURE[b]{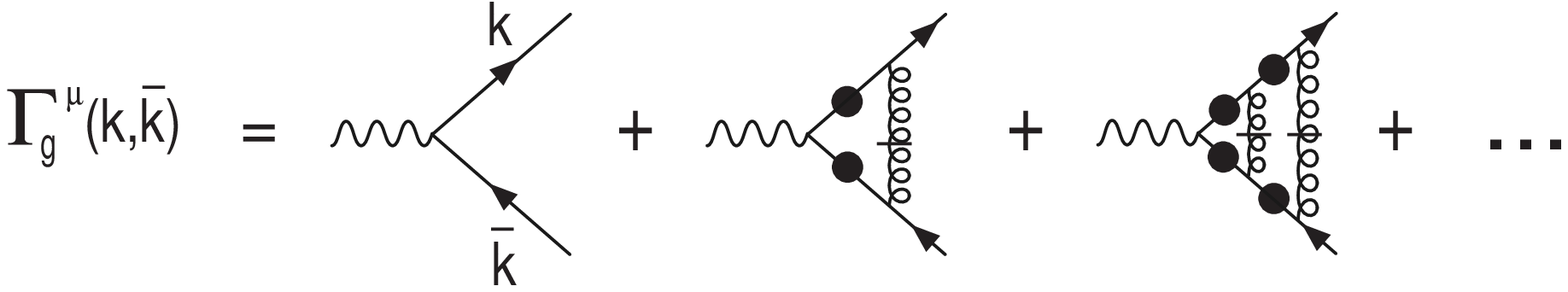,width=12cm}{Quark-photon vertex 
$\Gamma_g^{\mu}(k, \bar k)$ in a gluon condensate.}
\EPSFIGURE{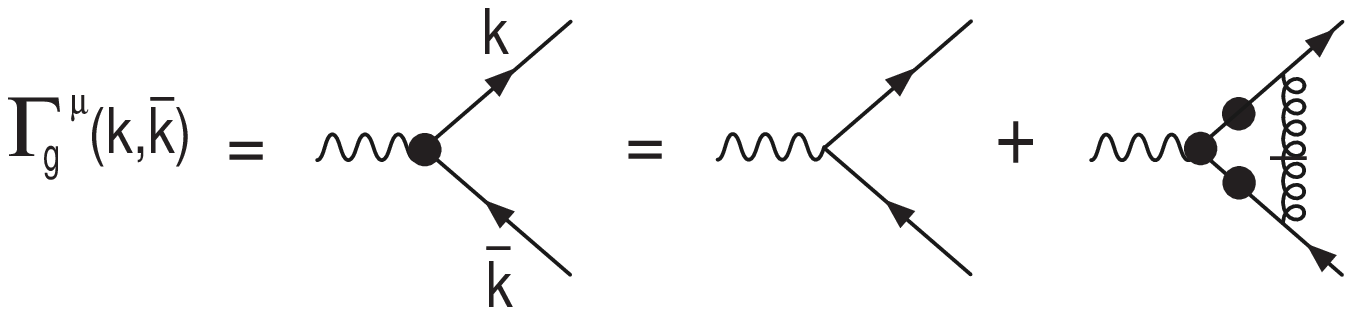,width=10cm}{Implicit equation for
$\Gamma_g^{\mu}(k, \bar k)$.}
When the vertex is expanded on its independent Dirac components this
equation reduces to a set of linear equations for the components. We
do this explicitly in Appendix C for the case $S_g = S_{g1}$ (\cf\
Eqs.~\eq{generalform} and \eq{lineqs}). Thus the solution is unique
when $S_g$ is given.

Multiplying \eq{Gammagmuimplicit} by $p_\mu = (k-\bar k)_\mu$ and
noting the propagator identity \eq{Sg0},
\beq
\halft\mu_g^2\gamma^\mu S_g(k)\gamma_\mu = S_g^{-1}(k) - \Slash{k}
\eeq
one easily verifies that $p_\mu\Gamma_g^\mu = S_g^{-1}(k)-
S_g^{-1}(\bar k)$ is a solution. Hence we conclude that the unique
solution of \eq{Gammagmuimplicit} respects the Ward-Takahashi
identity \eq{WT}.

For highly virtual momenta $k^2,\ \bar k^2 \to \infty$, we have
$S_g(k) \to 1/\Slash{k}$, $S_g(\bar k) \to 1/\Slash{\bar k}$
and \eq{Gammagmuimplicit} thus implies
\beq \label{asp2g}
k^2,\ \bar k^2 \to \infty \ \ \Rightarrow \ \
\Gamma_g^{\mu}(k, \bar k) =
\gamma^\mu +\morder{\frac{\mu_g^2}{k^2}, \frac{\mu_g^2}{\bar k^2}}
\eeq
This may also be verified from the explicit expression
\eq{Gammagmuexplicit0} of the dressed vertex.

\subsection{Photon self-energy $\Pi_g^{\mu\nu}(p)$}

Given the novel analytic structure of the dressed
($\lambda_g \neq 0$) quark propagator (\cf\ \eq{csbcons} and
\eq{csbbreak}) it
is of interest to verify that the physical $p^2=0$ pole remains in
the photon propagator. At large $N$ a single (dressed) quark loop
dominates the photon self-energy $\Pi_g^{\mu\nu}(p)$.
Thus (\cf\ Fig.~10),
\beq
\label{selfg}
\Pi_g^{\mu\nu}(p) = ie^2 N \int \frac{d^D k}{(2\pi)^D}
{\rm Tr} \left[\gamma^{\nu}S_g(k)\Gamma_g^{\mu}(k,\bar k)S_g(\bar k)\right]
\eeq
\EPSFIGURE[b]{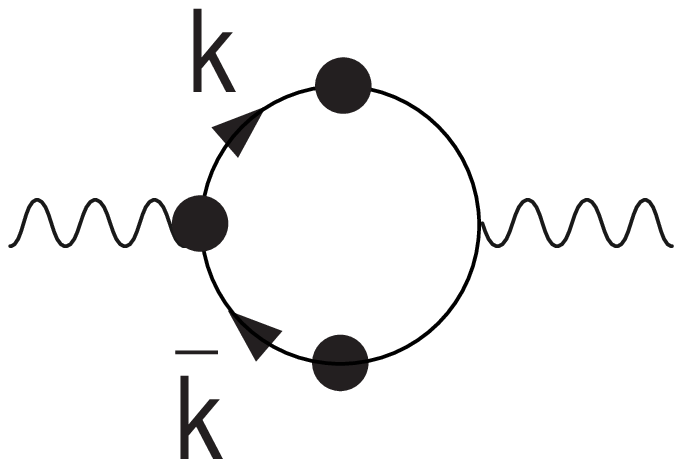,width=12cm}{Photon self-energy $\Pi_g^{\mu\nu}(p)$
in a gluon condensate.}
At highly virtual loop momenta the condensate dressing is power
suppressed, \cf\ Eqs.~\eq{asp2bis} and \eq{asp2g}. We dimensionally
regularize in $D$ dimensions
the ultraviolet quadratic and logarithmic divergences,
which  are independent of $\mu_g^2$.

Multiplying the self-energy \eq{selfg} by $p_\mu$ and using the
Ward-Takahashi identity \eq{WT} one finds $p_{\mu} \Pi_g^{\mu\nu}(p)
= 0$. Hence the self-energy is transverse,
\beq
\label{Pig}
\Pi_g^{\mu\nu}(p) = \Pi_g(p^2)\,(p^2 g^{\mu\nu} - p^{\mu}p^{\nu})
\eeq
In order to prove that the photon remains massless, we show that
$\Pi_{g \mu}^{\mu}(p) = 3p^2 \Pi_g(p^2)$ vanishes in the
$p^2 \to 0$ limit. Since this quantity only depends on $p^2$ we may
equivalently consider the 4-vector $p = k - \bar k \to 0$ limit.

Following Fradkin \cite{fradkin}, we note that the $k\to \bar k$
limit of the Ward-Takahashi identity \eq{WT} reads
\beq
\label{wardCgpto0}
p_{\mu} \Gamma_g^{\mu}(k, k)  = p_{\mu}
\frac{\partial S_g^{-1}(k)}{\partial k_{\mu}}
\eeq
Since the components $p^\mu$ are independent this yields the
Slavnov-Taylor identity,
\beq
\label{kkvertex}
\Gamma_g^{\mu}(k, k)  = \frac{\partial S_g^{-1}(k)}{\partial k_{\mu}}
=  - S_g^{-1}(k) \left[\frac{\partial S_g(k)}{\partial k_{\mu}}
\right] S_g^{-1}(k)
\eeq
Using \eq{kkvertex} in \eq{selfg} gives
\beq
\label{Atildemumu2}
\Pi_{g \mu}^{\mu}(p=0) =  -ie^2 N \int \frac{d^Dk}{(2\pi)^D}
\frac{\partial}{\partial k_{\mu}} \,
{\rm Tr} \left[\gamma_{\mu}S_g(k)\right] =0
\eeq
which vanishes within dimensional regularization
since the integrand is a full derivative. Hence the
resummed photon propagator
\beq
\frac{1}{p^2[1-\Pi_g(p^2)]}\left(-g^{\mu\nu} +
\frac{p^{\mu}p^{\nu}}{p^2}\right)
\eeq
has the physical photon pole at $p^2=0$.

We give an explicit integral expression for $\Pi_g(p^2)$ in
Eq.~\eq{Pimumu}, for the case of the quark propagator
$S_{g1}$ given in \eq{csbcons}.

\section{Outlook}

Our work raises several issues which merit further study. First of
all, we did not consider the condensate corrections to the gluon
propagator. The four-gluon coupling causes a richer set of
diagrams to contribute in this case. However, this appears to
be only a technical complication. The lowest order ($\ell=0, n=1$ in
\eq{pertexpg}) gluon condensate correction to the gluon  propagator was
evaluated in \cite{cr} and found to have the required transverse
structure. We have checked that the $n=2$ correction is also
transverse.
The expansion of the gluon self-energy dressed with a gluon condensate
starts as
\beq
\Pi^{\mu\nu}(p)=\Pi_0^{\mu\nu}(p)+\mu_g^2 \,
\left(g^{\mu\nu} - \frac{p^{\mu}p^{\nu}}{p^2}\right)
\, \left( -2 + \frac{21}{2} \frac{\mu_g^2}{p^2} + \ldots \right)
\eeq
where $\Pi_0^{\mu\nu}(p)$ is the standard perturbative ($\mu_g^2=0$)
result.

It will also be worthwhile to study the case where both $\lambda_g$ and
$\lambda_q$ are non-zero, which we have not considered here.
The richer set of diagrams then
appears not only for the gluon, but also for the quark
propagator.

We took the $N \to\infty$ limit to simplify the set of contributing
diagrams. The algebraic nature of the resummation holds for any $N$.
It would be interesting to derive the generalizations to finite $N$
of DS equations like \eq{Sq0} and \eq{Sg0}.

We considered only the case of a vanishing bare quark mass, \cf\ \eq{qpropmod}.
For $m_q \neq 0$ the
on-shell condition \eq{condterm} of a quark condensate with $\pvec=0$ would
have the form $\delta(p^0-m_q)$ and thus would break explicitly the Lorentz
symmetry. On the other hand, it is straightforward to generalize
our analysis to $m_q \neq 0$ in the presence of a gluon condensate
($\lambda_g \neq 0, \lambda_q = 0$), which preserves Lorentz invariance.
One finds that the modified implicit equation \eq{Sg0} for
the quark propagator $S_g(p)$ is then of fourth order in the parameters
$a,b$ of Eq.~\eq{genform}.

In this paper we did not
sufficiently address the treatment of
contributions proportional to the volume of space-time
$\delta^4(0)$. Such terms correspond to interactions between the
$\pvec=0$ condensate particles themselves and we expect that they
factorize from measurable quantities. A more systematic treatment of
these effects would be desirable.

The restriction to $\pvec=0$ condensates required for explicit
Lorentz invariance is very constraining. $C(\pvec=0)$ in the
asymptotic wave function \eq{psiC} is then the only parameter, which
translates into $\lambda_g$ and $\lambda_q$ in the free gluon and
quark propagators. We believe that systematic studies of the
properties of perturbative expansions with a $\pvec=0$ condensate
term are warranted.

Our result that the quark propagator dressed with a gluon
condensate \eq{csbcons} has a
cut and hence vanishes at large times (\cf\ \eq{ast})
addresses the question of the
analytic structure of confined partonic Green functions
\cite{Brower:id,Gribov:1999ui}. This obviously also has a profound
effect on unitarity relations -- as it should to describe a physical
situation where completeness sums involve hadrons rather than partons.

\acknowledgments
We thank T.~Binoth, S. Brodsky, A.~Cabo and E.~Pilon
for valuable discussions.
S.~P. would also like to thank the people of the SPhT in Saclay for
their interesting comments on our work.

\bigskip\bigskip
\centerline{APPENDIX}

\appendix

\section{Quark-photon vertex $\Gamma_q^{\mu}(k, \bar k)$:
$\lambda_g=0,\ \lambda_q \neq 0$.}
\label{sec:A}
The quark-photon vertex $\Gamma_q^{\mu}(k, \bar k)$ is given in terms
of the quantities $V^{\mu}$ and $W^{\mu}$, see \eq{Gammaqmudef}.
Here we solve the equations \eq{VWimplicit} for
$V^{\mu}$ and $W^{\mu}$. By iterating these equations
one can infer that $V^{\mu}$ and $W^{\mu}$ have the general form:
\beqa
\label{VWgeneral}
V^{\mu}(k, \bar k) &=& \gamma^{\mu} - \frac{\mu_q^3}{2 {\bar k}^2}
\left[A \gamma^{\mu} + B p^{\mu} + C \Slash{p} p^{\mu} \right] \nonumber \\
W^{\mu}(k, \bar k) &=& \gamma^{\mu} - \frac{\mu_q^3}{2 {k}^2}
\left[A' \gamma^{\mu} + B' p^{\mu} + C' \Slash{p} p^{\mu} \right]
\eeqa
where $A$, $B$, $C$, $A'$, $B'$, $C'$ are functions of $p^2$ only.
Writing the quark propagator \eq{Sq} as
\beq
\label{Sparam}
S_q(p) = a \Slash{p} + b
\eeq
\beq
a = \frac{p^4}{p^6 - (2\mu_q^3)^2}; \ \
b = \frac{-2 \mu_q^3 \, p^2}{p^6 - (2\mu_q^3)^2} \nonumber
\eeq
one then inserts \eq{VWgeneral} and \eq{Sparam} into \eq{VWimplicit}.
A straightforward calculation yields six equations for the unknown
quantities $A$, $B$, $C$, $A'$, $B'$, $C'$:
\beqa
A &=& - 2b \, ( 1 - \frac{\mu_q^3}{2 p^2} A')  \ \ ; \ \
B = 4a - \frac{2\mu_q^3}{p^2} \left(a A'+b B' +a p^2 C' \right) \ \ ;
\ \ C = \frac{\mu_q^3}{p^2} \left(a B'+b C'\right) \nonumber \\
A' &=& - 2b \, ( 1 - \frac{\mu_q^3}{2 p^2} A) \ \ ; \ \
-B' = 4a - \frac{2\mu_q^3}{p^2} \left(a A-b B +a p^2 C \right) \ \ ;
\ \ C' = \frac{\mu_q^3}{p^2} \left(-a B+b C\right) \nonumber \\
\eeqa
The solutions are found to be:
\beq
\label{ABC}
A = A' = \frac{4 \mu_q^3 p^2}{p^6 - 2 \mu_q^6} \ \ ;\ \
B = -B' = \frac{4}{p^2} \ \ ;\ \
C = C' = \frac{4 \mu_q^3}{2 \mu_q^6 -p^6}
\eeq
Using these expressions in \eq{VWgeneral} and \eq{Gammaqmudef} gives
the result \eq{Gammaqmu} for the dressed quark-photon vertex
$\Gamma_q^{\mu}(k, \bar k)$.

\section{Asymptotic behaviour of the dressed quark propagator
$S_g(t, \vec{p})$}
\label{sec:B}

We derive in this appendix the asymptotic time behaviour of the
dressed ($\lambda_g\neq  0,\ \lambda_q = 0$), chirally symmetric
quark propagator $S_{g1}(p)$ given in Eq.~\eq{csbcons}. This
propagator may be written as
\beq
\label{B1}
S_{g1}(p) = \frac{2 \Slash{p}}{\mu^2}
\left[ 1 - \frac{p^2-\mu^2}{\sqrt{p^2+\ieps}\sqrt{p^2-\mu^2+\ieps}}\right]
\eeq
where we define $\mu = 2\mu_g$ in the present appendix, and the $\ieps$
prescription arises from the usual Feynman prescription of the
free ($p\neq 0$) quark propagator $\Slash{p}/(p^2+\ieps)$.
The Fourier transformed propagator
\beq
S_g(t, \pvec) = \int_{-\infty}^{\infty} \frac{dp_0}{2\pi} S_g(p)
\exp(-itp_0)
\eeq
is thus
\beqa
S_g(t, \pvec) &=&
\frac{2 (i \gamma^0 \partial_t - \pvec\cdot\vec{\gamma} )}{\mu^2}
\left[\delta(t) + (\pvec^2 +  \mu^2 + \partial_t^2) J(t, \pvec^2, \mu^2)
\right] \label{Sgt} \\
J(t, \pvec^2, \mu^2) &=&  \int_{-\infty}^{\infty} \frac{dp_0}{2\pi}
\frac{\exp(-itp_0)}{\sqrt{p_0^2-\pvec^2+\ieps}\sqrt{p_0^2-\pvec^2-\mu^2+
        \ieps}} \label{J}
\eeqa
The function
$J(t, \pvec^2, \mu^2)$ can be evaluated using Feynman
parametrization,
\beq
\label{feynpar}
\frac{1}{\sqrt{A+\ieps}\sqrt{B+\ieps}} = \frac{1}{\pi}
\int_0^1\frac{dx}{\sqrt{x(1-x)}} \, \frac{1}{(1-x)A+xB+\ieps}
\eeq
and doing the $p_0$-integral using Cauchy's theorem.
The result is:
\beq
J(t, \pvec^2, \mu^2) = \frac{e^{-i|t\pvec|}}{2i\pi}
\int_0^1\frac{dx}{\sqrt{x(1-x)}}
\frac{1}{\sqrt{\pvec^2+x\mu^2}} \,
\exp\left[\frac{-i|t|x\mu^2}{\sqrt{\pvec^2+x\mu^2}+|\pvec|} \right]
\label{J2}
\eeq
The behaviour of $J(t, \pvec^2, \mu^2)$ for $|t| \to \infty$ can be inferred
by noticing that the integrand in \eq{J2} is peaked at $x \to 0$ in
this limit. With the change of variable $y=|t|\mu^2 x/(2|\pvec|)$ one
obtains
\beq
J(t, \pvec^2, \mu^2) \ \ \mathop{\longrightarrow}_{|t| \to \infty} \ \
\frac{e^{-i|t\pvec|}}{i\pi\sqrt{2|t\pvec|\mu^2}}
\int_0^{\infty} \frac{dy}{\sqrt{y}} e^{-iy} =
     - \frac{1+i}{2\sqrt{\pi}} \frac{e^{-i|t\pvec|}}{\sqrt{|t\pvec|\mu^2}}
\label{Jfinal}
\eeq
where we used
\beq
\int_0^{\infty}\frac{dy}{\sqrt{y}} \cos(y) =
\int_0^{\infty}\frac{dy}{\sqrt{y}} \sin(y) = \sqrt{\frac{\pi}{2}}
\eeq
Using \eq{Jfinal} in \eq{Sgt} gives the asymptotic time behaviour
\beq
S_g(t, \vec{p}) \ \ \mathop{\sim}_{|t| \to \infty} \ \
-\frac{1+i}{2\sqrt{\pi}} \,\ \frac{|\pvec| \gamma^0 -
      \pvec\cdot\vec{\gamma}}{\sqrt{|t\pvec|\mu_g^2}}
\, \exp(-i|t\pvec|)
\label{Sgasymp}
\eeq

\section{Quark-photon vertex $\Gamma_g^{\mu}(k, \bar k)$:
$\lambda_g\neq  0,\ \lambda_q = 0$.}
\label{sec:C}

In this appendix we solve the implicit equation \eq{Gammagmuimplicit}
for the $q\bar q\gamma$ vertex $\Gamma_g^{\mu}$ in the case of the
chirally invariant quark propagator \eq{csbcons},
\beq \label{ap}
S_{g1}(p) = a_p\, \Slash{p}\ \ ;\ \ a_p =
\frac{1}{2\mu_g^2}\left(1-\sqrt{1-\frac{4\mu_g^2}{p^2}}\right)
\eeq
Using this expression for $S_{g}$ Eq.~\eq{Gammagmuimplicit} becomes
\beq
\label{C2}
\Gamma_g^{\mu}(k, \bar k) =  \gamma^{\mu} - \halft f
\gamma^{\nu} \Slash{k} \Gamma_g^{\mu}(k, \bar k) \Slash{\bar k} \gamma_{\nu}
\eeq
where we introduced another dimensionful parameter $f$,
\beq
\label{f}
f = \mu_g^2 a_k a_{\bar k}
\eeq
Chiral and parity invariance restricts $\Gamma_g^{\mu}(k, \bar k)$ to the form
\beq
\label{generalform}
\Gamma_g^{\mu}(k, \bar k) =  A_0 \gamma^{\mu} + A_1 k^{\mu} \Slash{k}
+ A_2 k^{\mu} \Slash{\bar k} + A_3 {\bar k}^{\mu} \Slash{k}
+ A_4 {\bar k}^{\mu} \Slash{\bar k} + i A_5 \gamma_5
\epsilon^{\mu}(\gamma, k, \bar k)
\eeq
where we defined
\beq
\epsilon^{\mu}(\gamma, k, \bar k) = \epsilon^{\mu\nu\rho\sigma}
\gamma_{\nu} k_{\rho} {\bar k}_{\sigma}
\eeq
We find the coefficients $A_i$ by inserting \eq{generalform} into
\eq{C2} and using
\beqa
\Slash{\bar k} \gamma^{\mu} \Slash{k} &=& k^{\mu} \Slash{\bar k} +
{\bar k}^{\mu} \Slash{k}  - k\cdot{\bar k} \, \gamma^{\mu} -  i \gamma_5
\epsilon^{\mu}(\gamma, k, \bar k)  \nonumber \\
i \gamma_5 \Slash{\bar k} \epsilon^{\mu}(\gamma, k, \bar k) \Slash{k} &=&
- i \gamma_5 \epsilon^{\mu}(\gamma, k, \bar k)\, k\cdot{\bar k}
+ \gamma^{\mu} \left[k^2 {\bar k}^2 - (k\cdot{\bar k})^2 \right]
\nonumber \\ &&- k^{\mu} \left[ {\bar k}^2 \Slash{k} - k\cdot{\bar
        k}\, \Slash{\bar k}\right]
- {\bar k}^{\mu} \left[k^2 \Slash{{\bar k}} - k\cdot{\bar k} \,
      \Slash{k}\right]
\eeqa
This gives the conditions
\beqa \label{lineqs}
A_0 &=& 1 - f k\cdot{\bar k} \, (A_0 +k\cdot{\bar k} \, A_5)
+ f k^2{\bar k}^2 A_5 \nonumber \\
A_1 &=& f {\bar k}^2 (A_2 - A_5) \nonumber \\
A_2 &=& f (A_0 + k\cdot{\bar k}\, A_5 + k^2 A_1) \nonumber \\
A_3 &=& f (A_0 + k\cdot{\bar k}\, A_5 + {\bar k}^2 A_4) \nonumber \\
A_4 &=& f k^2 (A_3 - A_5) \nonumber \\
A_5 &=& - f (A_0 +k\cdot{\bar k} \, A_5)
\eeqa
with solutions
\beqa
A_0 = \frac{1 + f k\cdot{\bar k}}{1+2f k\cdot{\bar k}+f^2k^2{\bar
        k}^2} \ \ &;& \ \ A_5 = \frac{-f}{1+2f k\cdot{\bar k}+f^2k^2{\bar
        k}^2} \nonumber \\
k^2 A_1 = {\bar k}^2 A_4 = - \frac{2fk^2{\bar k}^2}{1-f^2k^2{\bar
        k}^2} A_5 \ \ &;& \ \ A_2 = A_3 = -\frac{1+f^2k^2{\bar
k}^2}{1-f^2k^2{\bar
        k}^2} A_5
\eeqa
The result for $\Gamma_g^{\mu}(k, \bar k)$ then follows from
\eq{generalform}:
\beqa
\Gamma_g^{\mu}(k, \bar k)&=&
\frac{1}{1+2f k\cdot{\bar k}+f^2k^2{\bar k}^2} \left\{
(1 + f k\cdot{\bar k}) \gamma^{\mu} - f i\gamma_5 \,
\epsilon^{\mu\nu\rho\sigma} \gamma_{\nu} k_{\rho} {\bar k}_{\sigma}
\phantom{\frac{f^2}{f^2}} \right. \nonumber \\
&+& \left. \frac{2f^2}{1-f^2k^2{\bar k}^2}
(k^{\mu} \Slash{k} {\bar k}^2 + {\bar k}^{\mu} \Slash{\bar k} k^2)
+\frac{f(1+f^2k^2{\bar k}^2)}{1-f^2k^2{\bar k}^2}
(k^{\mu} \Slash{\bar k} + {\bar k}^{\mu} \Slash{k}) \right\}
\label{Gammagmuexplicit0}
\eeqa

Given this expression for the vertex one can directly show that it
satisfies the Ward-Takahashi relation \eq{WT}. Straightforward
algebra yields:
\beq \label{WT1}
p_{\mu}\Gamma_g^{\mu}(k, \bar k) = \Slash{k}
\frac{1-f{\bar k}^2}{1-f^2k^2{\bar k}^2} - \Slash{\bar k}
\frac{1-f k^2}{1-f^2k^2{\bar k}^2}
\eeq
Using \eq{f} and
Eq.~\eq{condSg} (for $b=0$),
\beq \label{quad}
a_p - \frac{1}{p^2} = \mu_g^2 a_p^2
\eeq
one gets
\beq
\frac{1-f{\bar k}^2}{1-f^2k^2{\bar k}^2} = \frac{1}{k^2 a_k} \ \ ;\ \
\frac{1-f{k}^2}{1-f^2k^2{\bar k}^2} = \frac{1}{\bar k^2 a_{\bar k}}
\eeq
Substituting these expressions in \eq{WT1} gives the Ward-Takahashi
relation \eq{WT}.

The explicit expression \eq{Gammagmuexplicit0} for the vertex
function allows us to write the photon self-energy $\Pi_g(p^2)$
defined by \eq{selfg} and \eq{Pig} as
\beqa
\label{Pimumu}
\Pi_g(p^2) = \frac{8 ie^2 N}{3p^2} \int \frac{d^Dk}{(2\pi)^D}
\left\{ \frac{\mu_g^2 k^2 {\bar k}^2 a_k^2 a_{\bar k}^2}{1 - \mu_g^4
k^2 {\bar k}^2
        a_k^2 a_{\bar k}^2} - \frac{k\cdot\bar k\,  a_k a_{\bar k} +
\mu_g^2 k^2 {\bar k}^2
       a_k^2 a_{\bar k}^2}{1 + 2 \mu_g^2 k\cdot\bar k
        a_k a_{\bar k} + \mu_g^4 k^2 {\bar k}^2 a_k^2 a_{\bar k}^2}
\right. \ \ &&
\nonumber \\
+ \left. \frac{\epsilon}{2} k\cdot\bar k\, a_k a_{\bar k}
\, \frac{1 +\mu_g^2 k\cdot\bar k a_k a_{\bar k}}{1 + 2 \mu_g^2 k\cdot\bar k
        a_k a_{\bar k} + \mu_g^4 k^2 {\bar k}^2 a_k^2 a_{\bar k}^2}
    \right\}\ \ \  &&
\eeqa

The factor $\epsilon$ in \eq{Pimumu} arises from the identity
$\gamma_{\mu}\gamma^{\mu}=D=4-\epsilon$. In the $p = k-\bar k \to 0$
limit the integrand reduces to
\beq
\label{Pimumu1}
\lim_{p\to 0} p^2\Pi_g(p^2) = \frac{2 ie^2 N}{3\mu_g^2} \int
\frac{d^Dk}{(2\pi)^D}
\left\{
\frac{\left(1-\sqrt{1-4\mu_g^2/k^2}\right)^3}{\sqrt{1-4\mu_g^2/k^2}}
- \epsilon \sqrt{1-4\mu_g^2/k^2} \right\} = 0
\eeq
The integral is found to vanish using Feynman parametrization as in
\eq{feynpar} and performing a standard Wick rotation. This confirms
the general result \eq{Atildemumu2}.


\end{document}